\documentclass[aps,prc,superscriptaddress,showpacs,floatfix,nofootinbib,notitlepage,twocolumn]{revtex4-2}
\usepackage{amsmath,graphicx,float,hyperref,upgreek,bbm}

\usepackage{xcolor}
\usepackage[normalem]{ulem}

\def\x{{\boldsymbol x}}
\def\r{{\boldsymbol r}}
\def\p{{\boldsymbol p}}

\def\y{{\boldsymbol y}}

\def\bra#1{\langle#1\vert}
\def\ket#1{\vert#1\rangle}

\newcommand{\rme}{{\mathrm e}}

\newcommand \beq{\begin{eqnarray}}
\newcommand \eeq{\end{eqnarray}}

\begin{document}
\title{Accessing zero-point fluctuations in high-energy nuclear collisions}
\author{Jean-Paul Blaizot}
\email{jean-paul.blaizot@ipht.fr}
\author{Jean-Yves Ollitrault}
\affiliation{Universit\'e Paris Saclay, CNRS, CEA, Institut de physique th\'eorique, 91191 Gif-sur-Yvette, France}

\begin{abstract}
  Ultracentral high-energy collisions between identical nuclei are very sensitive to the size and shape of the frozen configuration of nucleons at the instant of the collision, an information that is encoded in the ground state wave functions of the colliding nuclei. 
    The relative standard deviation of the size  of the colliding system soon after the collision has recently been inferred from event-by-event fluctuations of the momentum per particle in collisions of $^{208}$Pb nuclei at the LHC, and since a single collision event produces thousands of particles, this information is very precise.   
   We connect this extracted value to the magnitude of the zero-point fluctuations of the radius of the lead nucleus. The value  that is extracted from the data is compatible with estimates based on nuclear sum rules, but is overestimated by classical Glauber-type  modeling of  high-energy collisions. The role of the Pauli principle in suppressing the fluctuations in closed-shell nuclei is underlined. It leads to a distinctive $A$ dependence that could be checked in collisions of smaller systems. 
\end{abstract}

\maketitle

\section{Introduction}
A collision between two nuclei whose relative velocity is close to the speed of light creates a little  drop of quark-gluon plasma~\cite{Busza:2018rrf} which appears shortly after the nuclei have passed through each other~\cite{Heinz:2009xj}. This plasma expands freely into the vacuum, and eventually produces thousands of hadrons~\cite{ALICE:2016fbt}, a fraction of which are seen in detectors. 
The momenta and directions of these hadrons are determined by the size~\cite{Broniowski:2009fm} and the shape~\cite{Teaney:2010vd} of the quark-gluon plasma at the time of its creation,  reflecting the configuration of the nucleon positions at the instant of the collision~\cite{Giacalone:2019pca,Giacalone:2020awm,Jia:2021tzt}. Event by event, these configurations occur randomly according to the wave functions of the colliding nuclei, and they determine the overlap area between the projections of incoming nuclei onto the transverse plane, perpendicular to the collision axis~\cite{Ollitrault:2023wjk}. This entails a connection between observables accessible in relativistic heavy ion collisions and properties of nuclear ground states, a connection which has been very much explored over the recent years~\cite{Jia:2022ozr,Giacalone:2025vxa,Jia:2025wey}: 
Quadrupole deformation and triaxiality~\cite{Bally:2021qys,Jia:2021qyu,Bally:2023dxi,Giacalone:2024luz}, octupole~\cite{Zhang:2021kxj} and hexadecapole~\cite{Ryssens:2023fkv,Xu:2024bdh} deformations, neutron skin~\cite{Giacalone:2023cet}, nucleon-nucleon correlations~\cite{Zhang:2024vkh}, and $\alpha$ clustering~\cite{Broniowski:2013dia} in $^{16}$O~\cite{YuanyuanWang:2024sgp} and $^{20}$Ne~\cite{Liu:2025zsi} nuclei. 

Ultracentral collisions between identical spherical nuclei~\cite{Luzum:2012wu,CMS:2013bza}  which fully overlap, 
are a privileged observatory for nuclear structure~\cite{Liu:2022kvz,Kozyrev:2022ehy,Liu:2022xlm,Li:2025hae} because they offer essentially an image of a single nucleus~\cite{TabatabaeeMehr:2024lgu,Zhou:2025bwu}. 
Several studies suggest that correlations between outgoing particles can be mapped onto correlations in the ground state of the nucleus~\cite{Duguet:2025hwi,Blaizot:2025bfu,Mehrabpour:2026yuc}. 
In this paper, we argue that the zero-point fluctuation of the nuclear radius can be inferred from the event-by-event fluctuation of the transverse momentum per outgoing particle: 
For a fixed number of outgoing particles, i.e., a fixed multiplicity, a smaller quark-gluon plasma has a larger density, therefore a larger temperature and a larger momentum per particle. 
Size fluctuations  result therefore in fluctuations of the momentum per particle. 

These fluctuations have been analyzed at the LHC for Pb+Pb collisions with the same multiplicity, as a function of the multiplicity.  
The relative variance of the transverse size  can be precisely inferred~\cite{Alqahtani:2026edr} from these data~\cite{ATLAS:2024jvf}, as explained in the next section.  In the rest of this paper, we  show that the origin of these fluctuations, and their magnitude, can be traced back to the zero-point fluctuations of the nuclear ground state, and we recall how the latter can be estimated~\cite{Blaizot:1980tw}.
We explain why quantum effects reduce their magnitude  (as already pointed out by Zakharov~\cite{Zakharov:2021lux}) relative to a ``classical'' calculation where nucleons are sampled independently~\cite{Miller:2007ri}, as usually done when modeling nucleus-nucleus collisions. In closing this introduction, we note that the zero-point fluctuations of surface vibrations have been studied both in the context of  high-energy heavy-ion collisions~\cite{Hagino:2025vxe}, and that of low-energy fusion reactions~\cite{Esbensen:1981fk}. 

\section{Measuring event-by-event size fluctuations in heavy-ion collisions}
\label{s:qgp}

Our analysis will be based on the following available data: the transverse momenta of the produced particles and the total multiplicity, for each collision (event).  We denote by  $[p_T]$ the  
transverse momentum per particle in a given event:
\begin{equation}
  \label{defptperparticle}
[p_T]\equiv \frac{1}{N_{\text{ch}}}\sum_{j=1}^{N_{\text{ch}}} p_{T,j}\, ,
\end{equation}
where $p_{T,j}$ is the transverse momentum of an outgoing particle labelled $j$, and the sum runs over all particles seen in the event, whose number (the multiplicity of the event)  is denoted by $N_{\text{ch}}$. As we now argue, the knowledge of these two numbers allows us to infer global properties of the quark-gluon plasma produced in the collision, such as its size and temperature. 

We first recall that the standard theoretical description of nucleus-nucleus collisions at ultra-relativistic energies is based on relativistic hydrodynamics~\cite{Gale:2013da}. 
This provides a complete description of the expansion of the quark-gluon plasma, from the instant of its formation all the way to the freeze-out where it fragments into individual hadrons. 
We shall  assume that hydrodynamics starts at some early time, with no initial transverse velocity~\cite{Ollitrault:2007du}, and that the initial fluid is invariant under Lorentz boosts along the beam direction~\cite{Bjorken:1982qr}. 
Under these assumptions, the initial conditions are specified by the entropy density $s(\r,\tau_0)$ at a proper time $\tau_0$,  and the boost invariance implies that $s$ can  be viewed effectively as a two-dimensional field $s(\x)$ in the transverse plane (throughout the paper,  $\x$ denotes the coordinates in the transverse plane, i.e., $\r=(\x,z)$  with $z$ the longitudinal coordinate). Now, 
since the equations of hydrodynamics are deterministic, $N_{\text{ch}}$ and $[p_T]$ are entirely determined by $s(\x)$. 
This connection between the initial $s(\x)$ and the final observables $N_{\text{ch}}$ and $[p_T]$ is  usually obtained by solving numerically the equations of hydrodynamics~\cite{Gale:2013da}. 

Here we shall rely instead on an approximate description \cite{Gardim:2019xjs}, which turns out to be very accurate for our purpose, and  which allows us to relate directly $N_{\text{ch}}$ and $[p_T]$ to $s(\x)$.  
We  measure the total entropy\footnote{Strictly speaking, $S$ is the entropy per unit rapidity, divided by $\tau_0$.} $S$ and the transverse size $R$ by the first two non-trivial moments of $s(\x)$~\cite{Zhou:2025bwu}: 
\begin{align}
  \label{defS}
  S&\equiv \int_\x s(\x)\nonumber\\
  R^2&\equiv \frac{1}{S} \int_{\x}\x^2 s(\x). 
\end{align}
The origin is chosen at the center of mass of $s(\x)$, so that the first moment of $s(\x)$ vanishes,  $\int_\x \x \,s(\x)=0$.
It turns out that, to a good approximation, $N_{\text{ch}}$ and $[p_T]$ are completely determined by $S$ and $R$. 
On the one hand, $N_{\text{ch}}$ is proportional to the final entropy~\cite{Hanus:2019fnc}, which is itself proportional to $S$.\footnote{The entropy is conserved for an ideal fluid. We assume that its increase due to dissipation results in an overall multiplicative constant which does not depend on the details of initial conditions.}
On the other hand, hydrodynamic simulations show that $[p_T]$ is proportional to an effective temperature $T_{\rm eff}$ which mildly fluctuates event by event. In central Pb+Pb collisions at the LHC energy $\sqrt{s_{NN}}=5.02$~TeV, this effective temperature is approximately $T_{\rm eff}\approx 220$~MeV~\cite{Gardim:2019xjs,Gardim:2024zvi}. 
This effective temperature is related, through the equation of state, to the entropy density $s(T_{\rm eff})$, which is proportional to $S/R^3$ for dimensional reasons. 
Thus, hydrodynamics implies the relations~\cite{Alqahtani:2026edr}:  
\begin{align}
\label{effhydro}
N_{\text{ch}}&\propto S\nonumber\\
[p_T] &\propto T_{\rm eff}\nonumber\\
s(T_{\rm eff})&\propto \frac{S}{(R^2)^{3/2}},
\end{align}
where the proportionality factors are identical for all events. 

The underlying physical picture that eventually leads to these simple relations is rooted in the most generic features of the hydrodynamic evolution of a collision. This starts with a short period of time during which the system expands freely in the longitudinal direction, so that the total energy in a covolume,  $\varepsilon(\tau)\tau$, remains approximately constant. As time progresses, a longitudinal pressure builds up and thermalisation occurs. Finally, when the proper time becomes of the order of the transverse size of the system, $\tau\sim R$, the transverse expansion sets in. At that point the evolution is very fast, the pressure decreases rapidly so that the pressure work becomes negligible and both the energy and the entropy remain nearly constant until the freeze-out.  Thus  the effective temperature may be viewed as that of the system when the transverse expansion sets in. It is  determined entirely from the conservation of energy and entropy.    

The relations (\ref{effhydro}) imply that, at fixed $N_{\text{ch}}$, that is, at fixed $S$, a small variation of $R^2$ induces a small variation of $T_{\text{eff}}$ and hence of $[p_T]$~\cite{Broniowski:2009fm}.
Throughout this paper, $\delta f$ denotes the deviation of a quantity $f$ from its average value  $\langle f\rangle$ over events. 
 The variation $\delta [p_T]$ is obtained by differentiating Eqs.~(\ref{effhydro}), and by relating the variations of temperature and entropy density through the speed of sound $c_s^2=d\ln T/d\ln s$~\cite{Ollitrault:2007du}: 
\begin{equation}
  \left.\frac{\delta [p_T]}{\langle [p_T]\rangle}\right|_{\langle N_{\text{ch}}\rangle}=\left.\frac{\delta T_{\rm eff}}{\langle T_{\rm eff}\rangle}\right|_{N_{\text{ch}}}=-\frac{3}{2}c_s^2(T_{\rm eff})\left.
  \frac{\delta R^2}{\langle R^2\rangle }\right|_{N_{\text{ch}}}.
  \label{deltapt}
 \end{equation}
The ATLAS Collaboration has accurately measured the mean and the variance of $[p_T]$ in events with the same $N_{\text{ch}}$, as a function of $N_{\text{ch}}$~\cite{ATLAS:2024jvf}, in $^{208}$Pb+$^{208}$Pb collisions at $\sqrt{s_{NN}}=5.02$~TeV. 
Trivial statistical fluctuations due to the finite multiplicity are removed by subtracting the contribution from independent particles, so as to isolate the dynamical fluctuations that can be attributed to temperature fluctuations. 

An important source of dynamical fluctuations is that related to the event-by-event fluctuation of the impact parameter. 
The impact parameter $b$ defines the centrality of the collision.
As $b$ decreases towards   more central collisions, the overlap between the nuclei increases and more nucleons participate in the collision.
Since the multiplicity $N_{\text{ch}}$ is roughly proportional to the number of participant nucleons~\cite{Eremin:2003qn}, it is larger when $b$ is smaller.
However, this relation between $b$ and $N_{\text{ch}}$ holds only on average, and at a  fixed $N_{\text{ch}}$, there are fluctuations of $b$. For the largest values of $N_{\text{ch}}$ these fluctuations can be very accurately reconstructed via a data-driven and  model-independent analysis~\cite{Das:2017ned}.

The relevant fluctuations in the present discussion are those occurring at fixed $b$ and, more specifically, at $b=0$. 
This is the regime of the so-called ultracentral collisions   where the two spherical nuclei fully overlap.
Experimentally, ultracentral collisions are studied by selecting typically the $\approx 0.3\%$ fraction of collisions with the largest values of $N_{\text{ch}}$~\cite{CMS:2013bza,Das:2017ned}, in which $b$ becomes closer and closer to zero and fluctuates less. 
One can thus reconstruct the relative variance of $R^2$ from the data using Eq.~(\ref{deltapt}) if $c_s^2(T_{\rm eff})$ is known.
In practice, both $c_s^2(T_{\rm eff})$ and the variance of $R^2$ are reconstructed by fitting simultaneously data on the mean and variance of $[p_T]$ for large $N_{\text{ch}}$, using Eqs.~(\ref{effhydro}) and unfolding impact parameter fluctuations through a Bayesian analysis~\cite{Alqahtani:2026edr}. 
The speed of sound $c_s(T_{\rm eff})$ is then inferred from the increase of $\langle [p_T]\rangle$ with $N_{\text{ch}}$. The resulting value is in perfect agreement with first-principles calculations from lattice QCD~\cite{Borsanyi:2013bia,HotQCD:2014kol}, which lends support to the effective hydrodynamic description (\ref{effhydro}). 
Finally, the relative standard deviation of $R^2$ in Pb+Pb collisions at $b=0$ is found to be~\cite{Alqahtani:2026edr}\footnote{Note that this information, which is quite precise, is not accessible through standard global theory-to-data comparisons~\cite{Moreland:2018gsh,Nijs:2020roc,JETSCAPE:2020mzn}. 
The reason is that such analyses are calibrated without using data on ultracentral collisions. 
Consistent use of these data requires a precise evaluation of impact parameter fluctuations~\cite{Alqahtani:2026edr}, which depends on the experiment under study.} 
\begin{equation}
  \label{reverseengineering}
\xi_{\text{exp}}\equiv  \frac{\langle (\delta R^2)^2\rangle^{1/2}}{\langle R^2\rangle}= 3.09\pm 0.15\,\%  ,
\end{equation} 
where the error bar is evaluated by varying a number of parameters.\footnote{The two main contributions to the error are obtained by varying the range of the fit (lowest value of $N_{\text{ch}}$), and the parameters of the hydrodynamic calculation which is used to correct for the kinematic cuts of the detector (half of the particles, with low momenta, are not detected).} We have introduced here the notation $\xi$ to denote the relative standard deviation. This notation will be used throughout the paper, though not systematically.
The result (\ref{reverseengineering}) follows from the effective hydrodynamic description (\ref{effhydro}), supplemented by a key assumption that the entropy $S$ and the area $R^2$ are uncorrelated at $b=0$.
While this assumption is supported by general arguments~\cite{Zhou:2025bwu}, there may be a small residual correlation, whose effect is {\it not\/} taken into account in the error bar. 

In the rest of this paper, we shall argue that Eq.~(\ref{reverseengineering}) provides direct information on the zero-point fluctuations of the size of  the lead nucleus in its ground state.

\section{Fluctuations and correlations}
\label{s:fluctuations}

As we have indicated earlier, the initial conditions for the hydrodynamical evolution of the QGP are given in terms of a local random field $s(\x)$  in the transverse plane. Each event corresponds to a particular realisation of this random field. The correlations between the resulting fluctuations can be quantified in terms of the following correlation  function~\cite{Blaizot:2014nia}\footnote{We hope that no confusion will arise from our use of a conventional notation ${\cal S}$ for the density-density correlation function, and $S$ for the entropy.}
\begin{equation}
  \label{def2pointQGP}
{\cal S}(\x_1,\x_2)\equiv \langle \delta s(\x_1)\, \delta s(\x_2)\rangle, 
\end{equation}
in terms of which, for instance,  the variance of $R^2$ reads\footnote{We define the origin $\x=0$ as the center of the average density profile $\int_\x \x \,\langle s(\x)\rangle=0$. Throughout this paper, we neglect the recentering correction, i.e., the condition that $R^2$ is evaluated in a frame where $\int_\x \x \,s(\x)=0$ on an event basis, according to Eq.~(\ref{defS}), which typically decreases the variance by a factor $1-1/A$. }
\begin{equation}
  \label{varRQGP}
\langle (\delta R^2)^2\rangle=\frac{1}{S^2}\int_{\x_1,\x_2}
  \x_1^2 \x_2^2\, {\cal S}(\x_1,\x_2). 
\end{equation}
Our goal now is to relate this variance of the size of the quark-gluon plasma to the sources of the event-by-event fluctuations.
There are several sources of such fluctuations. 
Those associated with the particle production processes are local in $\x$ and are not correlated over distances larger than typically the nucleon size. 
The dominant fluctuations are those related to the positions of the nucleons at the instant of the collision~\cite{PHOBOS:2006dbo}. 
These can be correlated on distances that extend to the total size of the nucleus and are, in fact, \textit{the only ones that can exhibit such long-range correlations}.\footnote{These long-range correlations should not be confused with the short-range nucleon-nucleon correlations, which are often modelled by a minimum distance in Glauber calculations~\cite{Loizides:2017ack}.} 
These are the fluctuations that we are concerned with. 
They are directly related to the zero-point fluctuations present in the ground-state wave functions of the colliding nuclei. 
Such fluctuations are associated with the coherent, collective motions of all the nucleons, which manifest themselves more directly in collective excitations of the nucleus. 
Our task in this section will be to relate these zero-point fluctuations to those of the entropy field $s(\x)$. 

\subsection{The entropy density as a functional of the nuclear density}

The microscopic description of nucleus-nucleus collisions relies heavily on Glauber's theory of multiple scattering~\cite{Miller:2007ri}. 
In the optical limit, a crucial quantity that enters the theory is the thickness function, the integral of the nucleon density along the longitudinal axis: $t_A(\x)=\int_z \rho_A(\x,z)$, where $\rho_A(\r)$ is the one-body density of nucleons, normalised such that $\int_\r \rho(\r)=A$.\footnote{More precisely, $t_A(\x)$ involves a  convolution of the transverse density  with the density profile of a  single nucleon.
Eqs.~(\ref{symmetry}) and (\ref{2points2}) hold to linear order in the fluctuations and for a nucleon profile that is wide compared with the distance between nucleons in the transverse plane. 
Corrections for a narrower profile (see for instance the discussion in Appendix B of Ref.~\cite{Zhou:2025bwu}) are not evaluated here.}  
Event-by-event, the positions of the nucleons change, and the imprint of that change in the transverse plane is precisely captured by  $t_A(\x)$.\footnote{Note that because of the Lorentz contraction that affect both nuclei in the center of mass frame, the time it takes the two nuclei to cross each other is negligible, so that the positions of the nucleons in each nucleus can be considered as frozen during the collision. This statement depends on the frame and the picture would be different, for example, in the rest frame of one of the two colliding nuclei.} 

Consider now the collision of two nuclei, denoted  by $A$ and $B$. Particle production mechanisms suggest that the entropy density in the transverse plane is a functional of the thickness functions $t_A(\x)$ and $t_B(\x)$~\cite{Moreland:2014oya}. In fact both 
global theory-to-data comparisons~\cite{Bernhard:2016tnd,Nijs:2020roc,JETSCAPE:2020mzn}, as well as theoretical arguments~\cite{Zhou:2025bwu} that we shall revisit shortly, strongly favor an initial entropy density of the form 
\begin{equation}
  \label{trentop=0}
s(\x) \propto  \left(t_A(\x)t_B(\x)\right)^{1/2}. 
\end{equation}

It is instructive to review briefly the heuristic argument that leads to this result~\cite{Zhou:2025bwu}. A fairly general picture of the initial stage of a heavy ion collision assumes that the partons (mostly gluons)  that are freed initially have typical transverse momenta $p_\perp\lesssim Q_s$, where $Q_s$ is the saturation momentum \footnote{This is in particular what comes out of models based on a saturation picture, such as \cite{Lappi:2006hq,Eskola:1999fc}. Such models also implement automatically the boost invariance of the particle production processes}. Their contribution to the energy density is $\sim Q_s^4$. Since the energy density is produced on a time scale $\tau_s\sim 1/Q_s$, we have  $\varepsilon(\tau_s) \tau_s\sim Q_s^3$. On the other hand, in one nucleus, $Q_s^2$ is proportional to the density of gluons in the transverse plane, a quantity which, assuming additivity of the gluon distribution functions, scales approximately as the thickness function. It follows that $ Q_s^3\sim (t_A t_B)^{3/4}$. For some time the initial energy density decreases because of the longitudinal expansion, keeping  $\varepsilon(\tau)\tau$ constant, as argued earlier. Interactions eventually lead to thermalisation at some time $\tau_0\sim 1/T$. The temperature $T$ is related to the energy density at that time through $\varepsilon(\tau_0)\sim T^4$, implying $ T^3\sim \varepsilon(\tau_0)\tau_0 = \varepsilon(\tau_s)\tau_s \sim Q_s^3$.  
The entropy density is proportional to $T^3$, which implies $s(\tau_0)\tau_0= T^3/T=T^2=Q_s^2=(t_A t_B)^{1/2}$, which is Eq.~(\ref{trentop=0}) above. Note that the argument involves generic features of the evolution of the system,  as discussed above, and boost-invariant production mechanisms that are local in the transverse plane.

From the many-body perspective, the operator (\ref{trentop=0}) is a complicated operator. However in the case of two identical nuclei colliding at $b=0$, one can easily obtain the one- and two-point functions.  
For a central collision with $b=0$, linearising in the fluctuations, and omitting the global multiplicative constant, one obtains:
\begin{align}
\label{symmetry}
\langle s(\x)\rangle&=\langle t_A(\x)\rangle=\langle t_B(\x)\rangle\nonumber\\
\delta s(\x)&=\frac{1}{2}\left(\delta t_A(\x)+\delta t_B(\x)\right). 
\end{align}
The first relation identifies the average local entropy field with the thickness function to within a multiplicative constant. 
The two-point function (\ref{def2pointQGP}) contains four terms. 
Since the two nuclei are independent, the cross correlation vanishes:
$\langle\delta t_A(\x_1)\delta t_B(\x_2)\rangle=0$. 
The two other terms are equal by symmetry, and one obtains~\cite{Zhou:2025bwu}: 
\begin{equation}
  \label{2points2}
{\cal S}(\x_1,\x_2)=\frac{1}{2} \langle \delta t_A(\x_1)\delta t_A(\x_2)\rangle. 
\end{equation}
This equation allows us to identify the correlation function of the entropy density fluctuations to the corresponding correlation function of density fluctuations in the ground state of a single nucleus, to within  a multiplicative factor $\frac{1}{2}$. The latter quantity is amenable to a quantum evaluation. For greater clarity, we promote $t_A(\x)$ to a quantum operator in the transverse plane, and denote it with a hat, $t_A(\x)\mapsto \hat t_A(\x)$, and we substitute the average over events with a quantum average over the ground state many-body wave function and denote it as $\langle \cdots\rangle_0$.
With these notations, we write the two-body correlation function of a single nucleus as: 
\begin{equation}
\label{def2point0}
{\cal S}_A(\x_1,\x_2)\equiv\langle \hat t_A(\x_1)\hat t_A(\x_2)\rangle_0 -\langle \hat t_A(\x_1)\rangle_0\langle\hat t_A(\x_2)\rangle_0 .
\end{equation}
Similarly, we promote $R^2$ to a quantum operator,  $R^2\mapsto \hat R^2$ and, taking into account the relation~(\ref{symmetry}), we identify
\begin{equation}
  \label{twoversusone}
 \xi\equiv \frac{\langle (\delta R^2)^2\rangle^{1/2}}{\langle R^2\rangle}= \frac{1}{\sqrt{2}}\frac{\langle (\delta \hat R^2)^2\rangle_0^{1/2}}{\langle \hat R^2\rangle_0},
 \end{equation} 
 where the left-hand side represents the event-by-event average, while the right-hand side is the corresponding quantum average. 

\subsection{Calculating quantum averages}
\label{sec:2pointS}

To proceed further,  we have to specify the quantum averaging involved in the right-hand side of Eq.~(\ref{twoversusone}). Starting from the many-body wave function, and the associated  probability distribution of a nucleon configuration, $|\Psi_0(\r_1,\r_2,\cdots,\r_A)|^2$, we define  a reduced probability distribution in the transverse plane by integrating over the longitudinal coordinates
\beq
P(\x_1,\cdots, \x_A)=\int_{z_1,z_2,\cdots,z_A} |\Psi_0(\r_1,\r_2,\cdots,\r_A)|^2.
\eeq 
One can then define $n$-point functions in the usual way, by integrating over subsets of coordinates, e.g., 
\beq 
&&p^{(1)}(\x_1)=\int_{\x_2,\cdots,\x_A} P(\x_1,\cdots, \x_A),\nonumber\\
&&p^{(2)}(\x_1,\x_2)=\int_{\x_3,\cdots,\x_A} P(\x_1,\cdots, \x_A).
\eeq
Note that these $n$-point functions integrate to unity. In terms of these, we have $\langle \hat t_A(\x)\rangle_0 =A\,p^{(1)}(\x)$ and $\langle  \hat t_A(\x_1)  \hat t_A(\x_2) \rangle_0=A(A-1)\, p^{(2)}(\x_1,\x_2)$.
Consider now the operator $\hat R^2 =\frac{1}{A}\sum_i \x_i^2$ and its square $(\hat R^2)^2=\frac{1}{A^2}\sum_{ij} \x_i^2 \x_j^2$. Taking the expectation value with the probability defined above, one obtains
\beq
\langle (\hat R^2)^2\rangle_0 -\langle \hat R^2\rangle_0^2&=&\frac{1}{A^2} \sum_{ij} \left(\langle  \x_i^2 \x_j^2\rangle_0-\langle  \x_i^2 \rangle_0 \langle\x_j^2\rangle_0\right)\nonumber\\ 
&=& \frac{1}{A} \left(\langle \x^4\rangle_0 -\langle \x^2\rangle_0^2\right)\nonumber\\
&+&\left( 1-\frac{1}{A} \right) \left( \langle \x^2\y^2\rangle_0 -\langle \x^2\rangle_0\langle\y^2\rangle_0\right)\nonumber\\
\label{varianceofR2}
\eeq 
where $\langle \x^2\rangle_0 =\int_\x \x^2 \p^{(1)}(\x)$, and $\langle \x^2\y^2\rangle_0=\int_{\x,\y} \x^2 \y^2 p^{(2)}(\x,\y)$.

\subsection{Classical calculation}
\label{s:classical}

We conclude this section by presenting the result of a calculation that neglects all correlations in the many-body wave functions, essentially assuming that $|\Psi_0(\r_1,\r_2,\cdot,\r_A)|^2=\prod_ip^{(1)}(\x_i)$. 
In the absence of correlations between the particles, $p^{(2)}(\x,\y)=p^{(1)}(\x )p^{(1)}(\y)$, and Eq.~(\ref{varianceofR2}) reduces to the classical expression
\begin{equation}
  \label{varRPbclassical}
 \ \frac{\langle (\delta \hat R^2)^2\rangle_0^{1/2}}{\langle \hat R^2\rangle_0}
    =\frac{1}{\sqrt{A}}\left(\frac{\langle \x^4\rangle_0}{\langle \x^2\rangle_0^2}-1\right)^{1/2}. 
\end{equation}
This is typically the expression that is obtained in Glauber calculations, in which the probability of a given nucleon configuration is constructed by sampling the one-body density: This is indeed the only information that is needed to evaluate (\ref{varRPbclassical}).  
Note that the expression of the correlation function (\ref{def2point0}) leading to (\ref{varRPbclassical}) is:
\begin{equation}
  \label{def2point}
{\cal S}_A(\x_1,\x_2)\!=\! \langle \hat t_A(\x_1)\rangle_0 \delta(\x_1\!-\!\x_2)\!-\!\frac{1}{A}\langle \hat t_A(\x_1)\rangle_0\langle \hat t_A(\x_2)\rangle_0, 
\end{equation}
which satisfies the sum rule 
\begin{equation}
\label{sumrule}
\int_{\x_2}{\cal S}_{A}(\x_1,\x_2)=0, 
\end{equation}
reflecting particle-number conservation, or equivalently entropy conservation.\footnote{The subscript $A$ in ${\cal S}_{{A}}$ emphasises that this is the correlation function of a single nucleus, to be distinguished from that of the entropy fluctuations in Eq.~(\ref{def2pointQGP}). The two quantities differ by a factor $1/2$, according to Eq.~(\ref{2points2}).}
The first term represents  the correlation  associated with the local fluctuations of $\hat t_A(\x)$, and leads to the contribution $\langle \x^4\rangle_0$ in (\ref{varRPbclassical}),  while the second term enforces the sum rule~(\ref{sumrule}). 

The relative standard deviation given by (\ref{varRPbclassical}) scales with $A$ like $1/\sqrt{A}$, as expected for $A$ independent random variables (the nucleon positions).\footnote{Note that $\langle \x^4\rangle_0$ and $\langle \x^2\rangle_0$ contain an additional $A$ dependence, which cancels in the ratio $\langle \x^4\rangle_0/\langle \x^2\rangle_0^2$ of Eq.~(\ref{varRPbclassical}), and which is related to the fact that the nucleon density inside nuclei is roughly constant, so that the radius of nuclei scale as $A^{1/3}$. Such an effect is automatically taken into account in Glauber calculations that sample the one-body density.}
For $^{208}$Pb, using a standard Woods-Saxon parametrisation~\cite{Alver:2008aq} of the average nuclear density, $\rho(\r)=\rho_0/(\rme^{(r-R)/a}+1)$, with $R=6.62$~fm and $a=0.546$~fm,  we obtain  from Eq.~(\ref{twoversusone}) 
the following value of the relative standard deviation:
\begin{equation}
  \label{woodssaxontransverse}
 \xi_{\text{cl}}= \frac{\langle (\delta R^2)^2\rangle^{1/2} }{\langle R^2\rangle}= 3.61\%. 
\end{equation}
This is larger than the experimental result (\ref{reverseengineering}).
Furthermore, it represents a lower bound on the result that would be obtained in a full simulation based on a Glauber calculation:
One must include additional fluctuations, modeling quantum fluctuations inherent to the collision process, in order to explain the relative standard deviation of the entropy $S$ at $b=0$, which is found to be $2.77\pm 0.05\%$~\cite{Alqahtani:2026edr}.\footnote{Note that within the approximations made (Eqs.~(\ref{2points2}), (\ref{def2point0})), fluctuations of $S$ would vanish in the absence of additional fluctuations, according to the sum rule (\ref{sumrule}).} 
Tuning a TRENTo initial-state calculation~\cite{Moreland:2014oya} to reproduce this value, as done in Appendix~C of~\cite{Alqahtani:2026edr}, raises the number in Eq.~(\ref{woodssaxontransverse}) from 3.6\% to 4.3\%, larger by $39\%$ than the experimental result. 
This shows that a classical calculation significantly overestimates zero-point fluctuations, and motivates a fully-quantum calculation. 

\section{Zero-point fluctuation of the nuclear ground state}
\label{s:nucleus}

\subsection{Fluctuations and sum rules}

The most direct way to access the quantum fluctuations is via sum rules. 
Sum-rule arguments allow us to get a simple, yet fully quantum, estimate of the zero-point fluctuations. Consider the  one-body monopole operator $M=\sum_{i=1}^A \r_i^2$.  Its  
 fluctuation (or variance)  is given by the general formula
\beq
\label{m0}
\langle \Delta M^2\rangle\equiv\bra{0} M^2 \ket{0} -\bra{0} M\ket{0}^2=\sum_{n\ne 0} \left| \bra{0} M \ket{n} \right|^2,
\eeq
where $\ket{n}$ denotes  the excited states of the nucleus that can be reached from the ground state $\ket{0}$ by the action of $M$. In general $m_0(M)\equiv\sum_{n\ne 0} \left| \bra{0} M \ket{n} \right|^2$ is not known. However, consider the following sum, where the transition matrix elements are weighted by the excitation energy $E_n-E_0$, 
\beq
\label{m1}
m_1(M)&=&\sum_{n\ne 0} (E_n-E_0)\bra{0} M \ket{n} \bra{n} M \ket{0}\nonumber\\
&=& \frac{1}{2} \bra{0} \left[ M,\left[H,M\right]\right]\ket{0}.
\eeq
When the interaction part of the nuclear hamiltonian contains only local forces, the double commutator involves only the kinetic energy, and can therefore be calculated explicitly, leading to the so-called Energy-Weighted Sum Rule (EWSR)~\cite{Bohigas:1978qu}
\beq
\label{EWSR}
m_1(M)= 2\frac{\hbar^2}{m} A\langle \r^2\rangle_0,
\eeq
where $m$ is the nucleon mass and $\langle \r^2\rangle_0$ the rms radius of the nucleus.  It follows that when the EWSR is exhausted by a single state, which to a good approximation is the case for the giant monopole resonance in heavy nuclei, Eq.~(\ref{m1}) gives $m_1(M)\approx E_0^+ m_0(M)$, with $E_{0}^+$ the energy of the resonance, and Eq.~(\ref{m0}) then gives 
\beq
\langle \Delta M^2\rangle=\frac{1}{E_{0}^+} m_1(M).
\eeq 
Using Eq.~(\ref{EWSR}) and $\langle M\rangle= A\langle \r^2\rangle_0$, 
one obtains the following expression for the relative standard deviation of $M$: 
\beq\label{eq:DeltaQ2From SR}
 \frac{\sqrt{\langle \Delta M^2\rangle} }{ \langle M\rangle}= \sqrt{\frac{2\hbar^2}{m A\langle \r^2\rangle_0 }  \frac{1}{E_0^+ }  }, 
\eeq
where $E_0^+$ is the excitation energy of the giant monopole resonance. 
This expression can be estimated in terms of experimental data. For the energy of the Isoscalar Giant Monopole Resonance (GMR) we take the  value quoted in~\cite{Patel:2013uyt}, namely  $E_0^+ A^{1/3} =81.2\pm 0.6$~MeV. Now, specifying to $^{208}$Pb, and 
taking $\langle \r^2\rangle_0\simeq\frac{3}{5} r_0^2 A^{2/3} \simeq (5.51)^2 {\text{fm}}^2$ (using $r_0=1.2$ fm) and $E_0^+=81.2 A^{-1/3}\simeq 13.7$~MeV, we get $\sqrt{\langle \Delta M^2 \rangle}/\langle M\rangle\simeq 1.087 A^{-2/3}\simeq 0.031$, a value compatible with an early estimate of this quantity (see~\cite{Blaizot:1980tw}). Note that this estimate uses for $\langle \r^2\rangle_0$ the value of the rms charge radius~\cite{Sun:2025qll}, so that the value $0.031$ could be considered as an upper bound.

Before we can compare these estimates with experimental data, two corrections need to be implemented. The first one is trivial and amounts to a simple multiplication by a factor $1/\sqrt{2}$, according to Eq.~(\ref{twoversusone}). The second correction is more subtle and involves the projection of the three-dimensional fluctuations of the nucleus onto the transverse plane. The fluctuations that we are interested in are indeed those of $\x^2=\sum_{i=1}^A (x_i^2+y_i^2)$, an operator that contains  monopole and  quadrupole components. Let us set $Q=\sum_{i=1}^A(2z_i^2-x_i^2-y_i^2)/2$, then  $\sum_{i=1}^A (x_i^2+y_i^2)= \frac{2}{3} (M-Q)$. It is easily seen that the variance of $\x^2$ is $\langle (\Delta\x^2)^2\rangle_0=\frac{4}{9}(\langle\Delta M^2\rangle +\langle\Delta Q^2\rangle)$ (the covariance of $M$ and  $Q$ vanishes by rotational symmetry for a spherical nucleus). The sum-rule argument used before for the giant monopole resonance can be used also to estimate the variance of $Q$. One finds $m_1(Q)=\frac{1}{2} m_1(M)$, so that $\langle\Delta Q^2\rangle_0=\frac{\hbar^2}{m} \frac{A\langle \r^2\rangle_0}{E_2^+}$, where $E_2^+$ is the energy of the giant quadrupole resonance (GQR). Finally, reinstating a factor $1/A$ in the definition of $\x^2$ in order to match with the notation of the previous section, i.e. $\x^2\mapsto \frac{1}{A}\sum_{i=1}^A (x_i^2+y_i^2)$, we get
\beq
\langle \Delta (\x^2)^2\rangle_0=\frac{4}{9}\frac{\hbar^2}{m A} \langle\r^2\rangle_0 \left(\frac{2}{E_0^+}+\frac{1}{E_2^+}  \right),
\eeq
and since $\langle \x^2\rangle_0 =\frac{2}{3} \langle \r^2 \rangle_0$, we have
\beq\label{xiquE0E2}
\frac{\sqrt{\langle \Delta (\x^2)^2\rangle_0}}{\langle \x^2\rangle_0}=\sqrt{ \frac{\hbar^2}{m A \langle\r^2\rangle_0 } \left(\frac{2}{E_0^+}+\frac{1}{E_2^+}  \right)   }
\eeq
Note that when $E_0^+=E_2^+$, this formula reduces to Eq.~(\ref{eq:DeltaQ2From SR}) in which we substitute $\langle\r^2\rangle\mapsto \langle \x^2\rangle =\frac{2}{3} \langle \r^2\rangle$. Thus, although the quadrupole fluctuations are of comparable magnitude as the monopole ones, because $E_2^+\simeq 63 A^{-1/3}\simeq 10.6$~MeV is not too different from $E_0^+\simeq 13.7$~MeV, they contribute to just a correction. 
Quantitatively, one gets a correction factor $\sqrt{ \frac{1}{3} \left( 2+\frac{E_0^+}{E_2^+}\right) }\simeq  1.047$. This value is in fact an underestimate. 
It relies on the assumption that the GQR exhausts the EWSR in $^{208}$Pb. 
This is however not the case: about 15\% of the EWSR is carried by a low lying $2^+$ state at 4.09~MeV~\cite{Harakeh:1979fdv}. For a simple estimate of this contribution, we may substitute $\frac{1}{E_2^+}\mapsto \frac{f}{4.09}+\frac{1-f}{10.9} $ with $f=0.15$. The correction factor  changes then from 1.047 to 1.091.

Applying all these corrections to the relative standard deviation  $\sqrt{\langle \Delta M^2 \rangle}/\langle M\rangle\simeq 0.031$ obtained above, namely $0.031\mapsto 0.031\frac{1}{\sqrt{2}} \sqrt{\frac{3}{2}}\times 1.091= 0.029$, one finally gets the quantum result for $^{208}$Pb, 
\beq
\xi_{\text{qu}}=2.9\%
\label{quantumresult}
\eeq
or more generally, $\xi_{\text{qu}} \simeq 1.02\, A^{-2/3}$.

Several comments are in order: i) this result is nearly compatible with the value extracted from data  (\ref{reverseengineering}),\footnote{Note that the result being somewhat smaller than the experimental value (slightly above the estimated error bar) leaves some room for other fluctuations associated with the process of particle production, as discussed at the end of Sec.~\ref{s:classical}.} 
ii) it is smaller than the classical estimate (\ref{woodssaxontransverse}), 
and iii) it has a different $A$ dependence than the classical result (\ref{varRPbclassical}): $A^{-2/3}$ versus $A^{-1/2}$. 
The last two points reflect the quantum suppression of the fluctuations, which to a large extent, can be attributed to the Pauli principle. 
The relevance of this quantum suppression for heavy-ion phenomenology has already been pointed out by Zakharov~\cite{Zakharov:2020irp,Zakharov:2021lux}, who has shown that the suppression of quadrupole fluctuations provides a natural explanation for the suppression of elliptic flow observed in ultracentral collisions (the so-called ``ultracentral flow puzzle''~\cite{Carzon:2020xwp,Giannini:2022bkn,Kuroki:2023ebq,Xu:2025cgx}). 

\subsection{Quantum suppression of zero-point fluctuations}
\label{s:pauli}

An important aspect of the breathing mode of heavy nuclei is that the interactions play a minor role: 
The compressibility of nuclear matter is close to that of a free Fermi gas~\cite{Blaizot:1980tw}. It is then appropriate to look at the correlations that make a Fermi gas distinct from a gas of independent classical particles

The density-density correlation function of a Fermi gas can  be expressed in terms of the one-body density matrix $\rho(\r_1,\r_2)$, whose non-diagonal elements capture the quantum effects induced by the antisymmetry of the many-body wave function. A standard calculation yields
 \begin{equation}
  \label{2pointquantum2}
  {\cal S}_A(\x_1,\x_2)=\rho(\x_1)\delta(\x_1-\x_2)-
  |\rho(\x_1,\x_2)|^2,
\end{equation}
where $|\rho(\x_1,\x_2)|^2=\int_{z_1,z_2} |\rho(\r_1,\r_2)|^2$ (see Section~\ref{sec:2pointS}).
The one-body density matrix can itself be written as  a projector on occupied single-particle states, $\rho(\x_1,\x_2)=\bra{\x_1}\rho\ket{\x_2}=\sum_h \int_{z_1,z_2} h(\x_1,z_1) h^*(\x_2,z_2)$, where $\sum_h$ runs over the occupied orbitals $h(\r)$. By using Eq.~(\ref{2pointquantum2}) to calculate the variance, we get then
\beq\label{eq:DeltaR2ph}
\langle (\delta \hat R^2)^2 \rangle_0=\frac{1}{A^2}\left[\langle \x^4\rangle_0 -\sum_{h h'}\bra{h}\x^2\ket{h'}\bra{h'} \x^2\ket{h}\right]
\eeq
where we have used $\sum_h \bra{h} \x^4 \ket{h}=\langle \x^4 \rangle_0$. 
By using the completeness relation in the space of single-particle states, 
$\mathbbm{1}= \sum_h |h\rangle \langle h|+\sum_p |p\rangle \langle p|$ 
where $|p\rangle$ denote an unoccupied orbital, one can rewrite Eq.~(\ref{eq:DeltaR2ph}) in the form: 
 \begin{equation}
 \label{particlehole}
\langle (\delta \hat R^2)^2\rangle_0=\frac{1}{A^2}\sum_{h,p}\left|\langle h|\x^2|p\rangle\right|^2. 
\end{equation}
This result could have been obtained directly from the sum rule, taking into account the fact that in the absence of interactions, the excited states are elementary particle-hole excitations. The detour by the correlation function was made in order to  illustrate the origin of the cancellation of the local term $\langle \x^4 \rangle_0$ in Eq.~(\ref{eq:DeltaR2ph}). This cancellation may be seen as the origin of the different $A$ behaviours of the quantum and classical results. 

Now, an operator such as $\x^2$ has significant particle-hole matrix elements for a limited set of orbitals around the Fermi surface (in the oscillator model to be discussed in the next section, $\x^2$  connects only  states that are separated by $2\hbar\omega$). Because of this, we expect a growing quantum suppression of the fluctuations with respect to the classical ones, in line with the different dependence on $A$ that has been identified in the respective cases.

\subsection{Simple estimates within the oscillator model}

Some of the previous considerations are nicely illustrated in a simple model of light nuclei, in which we assume that the nucleons occupy the lowest orbitals of a spherical harmonic oscillator. This model gives a fair, semi-quantitative, description of light nuclei, and it provides a nice illustration of the role of the Pauli principle. We shall analyze in some details results obtained for two closed-shell nuclei, namely $^{16}$O and $^{40}$Ca. But before doing so, we  consider the model for not so small nuclei,  in order to illustrate the emergence of the specific $A$ dependence of the quantum fluctuations of closed-shell nuclei.

Let us call $N$ the principal quantum number of  the last occupied shell. For not too small $N$ the number of nucleons in the nucleus $A\propto N^3$, while the degeneracy of the shell varies as $N^2$ \cite{blaizot1986quantum}. This degeneracy is essentially the number of particle-hole matrix elements involved in the calculation  of the fluctuation according to Eq.~(\ref{particlehole}). The size of the matrix element of the operator $\r^2$, which involves $2\hbar\omega$ transitions, is of order $N/\alpha^2$. Here $\alpha\equiv\sqrt{m\omega/\hbar}$, where  $m$ is the nucleon mass, and $\omega$ is the frequency of the oscillator, which scales as $A^{-1/3}$ to account for the fact that the rms radius of nuclei ($\sim 1/\omega$) grows as $A^{1/3}$. It follows that the sum over particle-hole matrix elements in (\ref{particlehole}) scales as $N^2\times N^2 \times \alpha^{-4}\sim A^{4/3}\times A^{2/3}\sim A^2$.  In contrast, the $A$-dependence of the classical variance of $\sum_i (x_i^2+y_i^2)$  is that of $A\langle \x^2 \rangle^2\sim A \times (A^{2/3})^2\sim A^{7/3}$, slightly larger than $A^2$ and in agreement with our previous findings.

Consider now explicitly $^{16}$O and $^{40}$Ca. In the case of $^{16}$O, the lowest $1s$ state, and the three $1p$ states of the oscillator are occupied by 4 nucleons each. In the case of $^{40}$Ca, in addition to the orbitals of $^{16}$O,  the orbitals $2s$ and the five orbitals $1d$ are occupied. In the case of   $^{16}$O all occupied orbitals can  be connected to an unoccupied one by the action of $\r^2$. However for $^{40}$Ca the $1s$ orbital is excluded, since the action of $\r^2$ would lead to another occupied orbital. In that case,  the effect of the Pauli principle should be more visible, as we shall indeed verify.  

The quantum calculation is easily performed in terms of sum rules. Since the GMR and the GQR resonances have the same energy $2\hbar\omega$, we have simply  $\langle (\Delta \x^2)^2\rangle_0 =\langle \x^2 \rangle_0 /{\alpha^2}$, that is, the variance of $\x^2$  is proportional to the expectation value of $\x^2$. The classical calculation involves the expectation value of $\x^4$, according to Eq.~(\ref{varRPbclassical}).  The expectation values $\langle \x^2 \rangle_0$ and $\langle \x^4 \rangle_0$ needed for estimating $\xi$ are given in Table~\ref{tab:results}. 
\begin{table}
    \centering
    \begin{tabular}{|c|c|c|c|c|}\hline
         & $ \alpha^2 A \langle \x^2\rangle_0 $  & $\alpha^4 A\langle \x^4 \rangle_0$  & $\xi^{\text{qu}}$ & $\xi^{\text{cl}}$ \\
         \hline
       $^{16}$O  & $24$ & $64 $ & $\frac{1}{4\sqrt{3}}\simeq 0.144$ &$\frac{1}{12}\sqrt{\frac{7}{2}}\simeq 0.156$ \\
       \hline
      $^{40}$Ca   & $80 $ & $272$ & $\frac{1}{4\sqrt{10}}\simeq 0.079$ &$\frac{1}{20}\sqrt{\frac{7}{2}}\simeq 0.093$ \\
      \hline
    \end{tabular}
    \caption{Values of some expectation values in the ground state of $^{16}$O and $^{40}$Ca in the simple harmonic oscillator model of light nuclei. The parameter $\alpha=\sqrt{m\omega/\hbar}$ has the dimension of an inverse length. The last two columns give the values of $\xi$ as obtained respectively from a quantum (see (\ref{xiquE0E2}) with $E_0^+=E_2^+=2\hbar\omega$) or a classical (see (\ref{varRPbclassical})) calculation. }
    \label{tab:results}
\end{table}
Note that the estimates of $\xi$ given in the table include the factor $1/\sqrt{2}$ implied by the mapping (\ref{twoversusone}). For instance, $\xi_{\text{O}}^{\text{qu}}=\sqrt{24}/(24\sqrt{2})\approx 0.144$.

By comparing the values of $\xi_{\text{O}}^{\text{qu}}$ and $\xi_{\text{O}}^{\text{cl}}$ one can see that even for  $^{16}$O where the Pauli blocking is ineffective, there is a (small) quantum suppression. This can be attributed to the cancellation of the term  $\langle\x^4\rangle$ which, as we have seen, leads to a slightly larger variance in the classical case. Even though the nuclei are too small to fully justify the general argument given above concerning the $A$ dependence of variances, it is interesting to look at the ratios 
\beq
\frac{\xi_{\text{Ca}}^{\text{qu}}}{\xi_{\text{O}}^{\text{qu}}}=\sqrt{\frac{3}{10}}\simeq 0.55,\qquad 
\frac{\xi_{\text{Ca}}^{\text{cl}}}{\xi_{\text{O}}^{\text{cl}}}=\frac{3}{5}= 0.6.
\eeq
By noting that $({16}/{40})^{2/3}\simeq 0.54$, while  $({16}/{40})^{1/2}\simeq 0.63$,  
we see that  the quantum calculation looks nicely compatible with the expected $A^{-2/3}$ law, while  the classical calculation ends up somewhat below the expected $A^{-1/2}$ law. Finally, to further gauge the effect of the quantum suppression, we may also take the ratio of the classical value of $\xi$ to its corresponding quantum value. We get
\beq
\left.\frac{\xi_{\text{cl}}}{\xi_{\text{qu}}}\right|_{\text{Ca}}=\sqrt{\frac{7}{5}}\simeq  1.18,\quad \left.\frac{\xi_{\text{cl}}}{\xi_{\text{qu}}}\right|_{\text{O}}=\sqrt{\frac{7}{6}}\simeq  1.08.
\eeq 
The ratio increases from  O to Ca, suggesting a larger suppression of the fluctuations in Ca than in  O, as anticipated. 

\section{Conclusions}
 
We have shown that the magnitude of transverse-momentum fluctuations measured  in ultra-central collisions at the LHC can provide quantitative information on the zero-point fluctuations of the size of the nuclei in their ground state. 
These fluctuations modify, event-by-event,  the transverse area of the collision zone determined by the positions of the nucleons at the instant of the collision. 
This changes in each event the size of the quark-gluon plasma produced in the collisions, eventually leading to measurable fluctuations in the distribution of the transverse momenta of the produced particles.  
The analysis of the available data on Pb+Pb  collisions at the LHC made it possible to reconstruct the relative standard deviation with an estimated uncertainty as small as 5\%. 
The quantum estimate is only slightly below the experimental value, leaving little room for small additional fluctuations related to the collision processes. 
On the other hand, the classical Glauber calculation overestimates this quantity by 16\% just from fluctuations in nucleon positions, and by 39\% if one adds fluctuations to match the observed multiplicity fluctuations. 
Thus data clearly favor a ground state with long-range correlations, resulting in a quantum suppression of fluctuations imputable to the Pauli principle. 
This quantum suppression modifies the $A$ dependence of the fluctuations, which could be checked by comparing $^{129}$Xe+$^{129}$Xe data~\cite{ATLAS:2024jvf} with $^{208}$Pb+$^{208}$Pb data. 
However, doing so requires a more  precise extraction of the fluctuations from the  present data, as well as a refined theoretical analysis to take into account the deformation of Xe. 
In any case, we find it quite remarkable that the  present LHC data, with their high statistics, allow us to address such questions at a quantitative level.

\bibliography{zeropoint}

\end{document}